\documentclass[aps,prl,showpacs,twocolumn,superscriptaddress]{revtex4}
 \usepackage{amsmath}
\usepackage{graphicx}
\usepackage{epsfig}
\newcommand{\beq}{\begin{equation}}
\newcommand{\eeq}{\end{equation}}
\newcommand{\bea}{\begin{eqnarray}}
\newcommand{\eea}{\end{eqnarray}}

\begin{document}
\bibliographystyle{apsrev}
 \title{ Nonequilibrium effective vector potential due to pseudospin exchange in graphene }
\author{  M. Kindermann}
\affiliation{ School of Physics, Georgia Institute of Technology, Atlanta, Georgia 30332, USA  }

\date{November 2007 }
\begin{abstract}
 We show that exchange interactions in two-dimensional electron gases out of equilibrium can generate a  fictitious vector potential with intriguing signatures in  interference and Hall measurements. Detailed predictions are made for graphene, where the effect is enhanced by pseudospin  exchange. 
       \end{abstract}
\maketitle


By driving a system out of equilibrium one typically  corrupts its quantum coherence. As a result many quantum effects observed in electronic structures, such as the Aharonov-Bohm interference and  the Kondo effect are suppressed as one applies a voltage bias  \cite{Ji:nat03,goldhaber:nat98,cronenwett:sci98}. In rare instances, however,  deviations from thermal equilibrium  modify quantum effects in more interesting ways, or  give rise to new coherent dynamics, such as  the AC-Josephson effect \cite{josephson:rmp74}. In this Letter we predict a novel effect in the latter category: a fictitious vector potential induced by nonequilibrium conditions. The effect is particularly robust and intuitive  in    graphene  \cite{novoselov:sci04,zhang:nat05,berger:jpc04}, where the conduction electrons  obey an equation that   resembles the relativistic Dirac equation \cite{novoselov:nma07}. In that equation an orbital degree of freedom  that is usually referred to as the `pseudospin' plays the role of the conventional electron spin   in the Dirac equation. This new quantum number has profound consequences, such as `Klein tunneling', allowing electrons to pass arbitrarily high potential barriers  \cite{katsnelson:nph06}, and a velocity renormalization   through exchange interactions \cite{barlas:prl07}. Here we show that in the nonequilibrium state created by a voltage bias the same pseudospin  exchange interactions induce an effective vector potential. Unlike the fictitious vector potentials created by lattice defects or distortions in graphene \cite{iordanskii:jet85,morozov:prl06,morpurgo:prl06} this effect does not cancel between the two bandstructure `valleys', but it has directly observable consequences. We discuss two of those signatures: i) interference currents oscillating as a function of the current density  in the material and ii) a `Hall' voltage that reveals  the fictitious magnetic field implied by the predicted  vector potential in curved conductors.

The origin of this effect is best understood within  the 
Dirac model for electrons in graphene at low energies,
\beq \label{Dirac}
H_\gamma= v \boldsymbol{\sigma_\gamma} \left( \boldsymbol{p}-\frac{e}{c}\boldsymbol{A }\right),
\eeq
where $\boldsymbol{ \sigma_\gamma}=(\sigma_x,\gamma\sigma_y)$ is a vector of Pauli matrices in pseudospin space,   $\boldsymbol{A }$  is the magnetic vector potential,   $\gamma=\pm $  the valley index,
 $\boldsymbol{p}$  the electron momentum, and $v$, $-e$, $c$ are the Fermi velocity, the electron charge, and the speed of light, respectively.  Eq.\ (\ref{Dirac}) implies that in each valley electrons  have a definite projection   of their pseudospin  onto their velocity.   When a bias voltage is applied that drives an electrical current, such that electrons have a preferred direction of motion, a graphene sheet therefore carries an excess pseudospin. Consequently one expects  a pseudospin-dependent exchange energy of electrons, described by an effective Hamiltonian with a vector potential $\boldsymbol{A^{\rm ex}}$ in addition to $\boldsymbol{A}$ in Eq.\ (\ref{Dirac}).

{\em Exchange vector potential:} To quantify the effect that is anticipated by the above simple Hartree-Fock argument  we first consider a clean, infinite sheet of extrinsic graphene, with Fermi energy $\varepsilon_{\rm F}>0$, at    $\boldsymbol{A}=0$. The effects of electron-electron interactions on single-particle quantities,  as considered here, are captured by the electron self-energy $\Sigma$. We evaluate $\Sigma$ in the so-called ${\rm G}_0{\rm W}$-approximation, resulting in a `screened   self-energy'. The `Fock'-contribution to the retarded self-energy then  is
\bea \label{G0W}
&& \Sigma_{\sigma\gamma}^{\rm R}(\varepsilon,\boldsymbol{p}) =  \frac{i}{2} \int \frac{d\omega}{2\pi}\frac{d^2\boldsymbol{k}}{(2\pi)^2}   \Big[ G_{\sigma\gamma }^{(0)\rm K}(\varepsilon-\omega,\boldsymbol{p}-\boldsymbol{k})  \\
&&\mbox{} \;\;\;\;\times W^{\rm R}(\omega,\boldsymbol{k} ) +G_{\sigma\gamma }^{(0) \rm R}(\varepsilon-\omega,\boldsymbol{p}-\boldsymbol{k}) W^{\rm K}(\omega,\boldsymbol{k})\Big]. \nonumber
\eea
Here, the noninteracting retarded and Keldysh \cite{Rammer:rmp86} Green functions $G^{(0)\rm R}$ and $G^{(0)\rm K}$, respectively, have an implicit matrix structure in pseudospin  space   and they are diagonal in the spin and valley indices $\sigma$ and $\gamma$. The retarded and the Keldysh components  of the RPA interaction potential $W^{\rm R}=v_{\rm C}/(1-v_{\rm C}\Pi^{\rm R})$ and $W^{\rm K}=\Pi^{\rm K} |W^{\rm R}|^2$, respectively, are found from  the unscreened Coulomb interaction $v_{\rm C}({\bf q})= 2\pi r_{\rm s} v/|\boldsymbol{ q}|$ at interaction parameter $r_{\rm s}$ and the polarizability 
$\Pi$. The bare Keldysh Green function $G^{(0)\rm K}$   is characterized by  occupation numbers $f_{s\boldsymbol{k}}$  of single-electron states $\psi_{\sigma\gamma s\boldsymbol{k}}$. Here, $\boldsymbol{k}$ are the electron momenta and $s= 1,-1$ for the electrons in the conduction band and the valence band, respectively. Through  $G^{(0)\rm K}$    also  $W^{\rm K}$ and $W^{\rm R}$ depend on $f_{s\boldsymbol{k}}$. We consider a   state with current flow in the direction ${\boldsymbol{\hat{w}_\theta}} =(\cos\theta,\sin\theta)$, $ f_{s\boldsymbol{k}}=\Theta[\varepsilon_{\rm F}+\Theta(s\boldsymbol{k}\cdot {\boldsymbol{\hat{w}_\theta}}) eV-\varepsilon_{s\boldsymbol{k}} ] $, with $\Theta(x)=1$ for $x>0$ and $\Theta(x)=0$ for $x\leq 0$. Here,  $\varepsilon_{s\boldsymbol{k}}$ are  the single-particle energies, $V$ is the voltage that drives the current, and we take the limit of temperature $T=0$, requiring $kT \ll eV$ in practice.   We assume weak nonequilibrium, $eV\ll \varepsilon_{\rm F}$, such that we may  expand in  small deviations $\delta f_{s\boldsymbol{k}} =V (\partial f_{s\boldsymbol{k}}/\partial V|_{V=0})$ of the occupation numbers from their equilibrium values. We find the leading order (in $eV/\varepsilon_{\rm F}$) nonequilibrium contribution to the self-energy    $\delta \Sigma_{\sigma\gamma}^{{\rm R}}= V (\partial\Sigma_{\sigma\gamma}^{{\rm R}}/\partial V|_{V=0})$ at small $r_{\rm s}$    by expanding the $G^{(0)\rm K}$ of Eq.\ (\ref{G0W}) in $\delta f_{s\boldsymbol{k}} $,
\bea \label{Fock}
\delta \Sigma_{\sigma\gamma}^{{\rm R}(0)}(\varepsilon,\boldsymbol{p}) &=& -\frac{1}{2 } \int\frac{d^2\boldsymbol{k}}{(2\pi)^2}   \sum_{s }\delta f_{s,\boldsymbol{k} }
\left(\begin{array}{cc}  1 & s e^{i \gamma \phi} \\ s e^{-i \gamma \phi} & 1 \end{array} \right) \nonumber \\
&& \mbox{} \;\;\;\;\;\;\;\;\;\;\;\;\; \;\;\;\;\; \;\;\;\times W^{\rm R}(\varepsilon-\varepsilon_{s\boldsymbol{k}},\boldsymbol{p}-\boldsymbol{k}) .
\eea
   Here,  $\boldsymbol{k}=k{\boldsymbol{\hat{w}_\phi}}$ and $W^{\rm R}$ takes its equilibrium form,  obtained in Refs.\ \cite{hwang:prb07,wunsch:njp06}. At  $eV \ll \varepsilon_{\rm F}$,   $\delta f_{s\boldsymbol{k}} $ is non-vanishing only if  $|\varepsilon_{s\boldsymbol{k}}- \varepsilon_{\rm F}|\ll \varepsilon_{\rm F}$. For all quantities computed below $\Sigma $ is needed, in the same limit, at   $|\varepsilon- \varepsilon_{\rm F}|\ll \varepsilon_{\rm F}$ and thus  $|\varepsilon-\varepsilon_{s\boldsymbol{k}}|\ll \varepsilon_{\rm F}$ in Eq.\ (\ref{Fock}). Since the typical  momentum transfer $\boldsymbol{p}-\boldsymbol{k}$ in Eq.\ (\ref{Fock}) is of order $k_{\rm F}$,   we may substitute $W^{\rm R}(\varepsilon-\varepsilon_{s\boldsymbol{k}},\boldsymbol{p}-\boldsymbol{k})  \approx W^{\rm R}(0,\boldsymbol{p}-\boldsymbol{k}) =  2\pi r_{\rm s} v/(|\boldsymbol{p}-\boldsymbol{k}|+q_{\rm sc})$ into Eq.\ (\ref{Fock}), with the screening wavevector $q_{\rm sc} = 4 r_{\rm s} k_{\rm F}$ at Fermi wavevector $k_{\rm F}$   \cite{hwang:prb07,wunsch:njp06}. The pseudospin-diagonal components of the real part  ${\rm Re}\, \delta \Sigma^{{\rm R}(0)}$   shift the chemical potential. The  off-diagonal components of the  hermitian part of $ \delta \Sigma^{{\rm R}(0)}$
     have a p-wave contribution that implies the velocity renormalization analyzed in Ref.\ \cite{barlas:prl07} and an s-wave part that acts as an effective vector potential $\boldsymbol{A^{\rm ex}}$,
   \beq \label{project}
\boldsymbol{A^{\rm ex}} = - \frac{c}{2ev} {\rm Re}  \int_0^{2\pi} \frac{d\vartheta}{2\pi}\,{\rm Tr}\, \boldsymbol{\sigma_\gamma} \delta \Sigma_{\sigma\gamma}^{{\rm R}(0)} (k_{\rm F}{\boldsymbol{\hat{w}_\vartheta}},\varepsilon_{\rm F})
\eeq
  (we set $\hbar=1$), which, for the above $f_{s\boldsymbol{k}} $, evaluates to
 \beq \label{Aex}
\boldsymbol{A^{\rm ex}} = \zeta \frac{c}{v} \frac{V}{8\pi}  {\boldsymbol{\hat{w}_\theta}},
\;\;
 \zeta =\int_0^{2 \pi}\frac{d\vartheta} {2\pi}\, \frac{4r_{\rm s}}{\sqrt{2-2\cos\vartheta}+4 r_{\rm s}}.
\eeq
 At  $r_{\rm s} \ll 1$ we have $\zeta \sim 4r_{\rm s} \ln(\pi/8r_{\rm s})/\pi$
and in the range of $0.15 < r_{\rm s}< 2.4$ typical for graphene \cite{DasSarma:07} $\zeta$ is of order unity ($\zeta\approx0.38$ at $r_{\rm s}=0.15$ and $\zeta\approx 0.89$ at $r_{\rm s}=2.4$). Note that the valley-dependence of the effect indeed drops out, as advertised \cite{footnote11}. %

Expanding  the $W^{\rm R}$ and the $W^{\rm K}$ of Eq.\ (\ref{G0W})  in $\delta f_{s\boldsymbol{k}}$ results in corrections to  $\zeta$. These corrections take the form of integrals over higher powers of $W^R$ than the leading contribution and   they are thus negligible at $r_{\rm s}\ll 1$, when the ${\rm G}_0{\rm W}$-approximation is reliable \cite{footnote2}. We therefore employ our 'screened Hartree-Fock' approximation  Eq.\  (\ref{Fock}) in the remainder of this Letter.  The Keldysh self-energy $\Sigma^{\rm K}$ and the anti-hermitian part of $\Sigma^{\rm R}$ account for inelastic relaxation. These inelastic processes are negligible in the geometries we study below: graphene ribbons with a finite length $L$ that are ballistic, that is  they have long inelastic and elastic mean free paths $ l_{\rm in}$ and $ l_{\rm el}$, respectively, $  l_{\rm in}, l_{\rm el} \gg L$, and  they are well-coupled to electron reservoirs  \cite{unpub}.  Neglecting relaxation by phonons ($T=0$), we find  from Eq.\ (\ref{G0W}) for a transport electron with momentum $\boldsymbol{k}=k{\boldsymbol{\hat{w}_\phi}}$  in the above nonequilibrium state  $f_{s\boldsymbol{k}} $ at $r_{\rm s}\ll 1$ and $eV \ll r_{\rm s} \varepsilon_{\rm F}$,
\beq  \label{lin}
l^{-1}_{\rm in} = k_{\rm F} \frac{\delta \varepsilon ^2(3eV-2 \delta\varepsilon )}{3\pi  \varepsilon^3_{\rm F}}\sum_{\eta=\pm 1} \left(\frac{r_{\rm s}}{|\theta-\phi+\eta \pi/2|+4 r_{\rm s}}\right)^{2}
\eeq
with $\delta \varepsilon=\varepsilon_{s\boldsymbol{k}}-\varepsilon_{\rm F}$, valid at $|\theta-\phi\pm \pi/2|\gg eV/\varepsilon_{\rm F}$  \cite{unpub}.

An effective vector potential $\boldsymbol{A^{\rm ex}}$  is generally expected in   electron systems where the exchange energy between two electrons depends on   their relative motion. This  requires    $q_{\rm sc} \lesssim k_{\rm F}$, which  is   typically  fulfilled in n-type GaAs, but not in p-type GaAs or Si-MOSFETs \cite{Sarma:prb05}. In graphene the effect persists into the strong screening limit $q_{\rm sc} \gg k_{\rm F}$ (at $r_{\rm s}\gg 1$) through pseudospin  exchange.
\begin{figure}\vspace{.5cm}
\includegraphics[width=8cm]{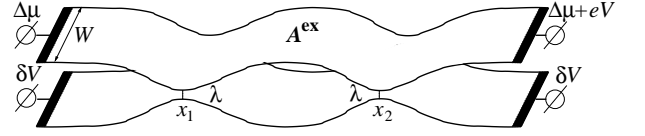}
\caption{ Two ballistic graphene ribbons attached to large leads (black bars), biased by a voltage $V$ such that a current flows through the upper ribbon.  Electrons tunnel between the ribbons at $x_1$ and $x_2$ with matrix element $\lambda$, driven by a voltage $\delta V$. The amplitudes for electron propagation between $x_1$ and $x_2$ along the upper ribbon with exchange vector potential $\boldsymbol{A^{\rm ex}}$  interfere with those along the lower ribbon, causing oscillations of the tunnel current as a function of $V$. 
 }  \label{fig1}
\end{figure}

{\em Interference:}   We now  turn to a  discussion of   signatures of the predicted effective vector  potential.
We first consider  two tunnel-coupled, ballistic graphene ribbons, as  shown in Fig.\ \ref{fig1}, where   $\boldsymbol{A^{\rm ex}}$ can be observed interferometrically \cite{footnote1}. The two ribbons are well-coupled to electron reservoirs at their ends and they touch at  $x_1$ and $x_2$, which creates a weak tunnel coupling $\lambda$ between them. For simplicity  we assume $\lambda$ to be  identical at $x_1$ and $x_2$ and constant over the width $W$ of the ribbons. We take the semiclassical limit $k_{\rm F} W \gg 1$ and  allow for different path lengths $ \Delta x^{\rm u}$ and  $ \Delta x^{\rm l}$ between the points of tunneling along the upper and the lower ribbon, respectively. A  voltage $V\ll \varepsilon_{\rm F}/e$ that leaves the mean electric potential invariant ($\Delta \mu =-eV/2+ \sqrt{\varepsilon_{\rm F}^2-(eV/2)^2}-\varepsilon_{\rm F}$ in Fig.\ \ref{fig1})  drives a current $I$  through the upper ribbon. According to Eq.\ (\ref{Aex}) an $\boldsymbol{A^{\rm ex}}$ results that shifts the phase of an electron traveling from $x_1$ to $x_2$  along the upper ribbon  by $\Phi= \zeta eV \Delta x^{\rm u}/8\pi v$.  The tunneling current $I_{\rm tun} $ between the ribbons, driven by a small voltage $\delta V $, is expected to oscillate as a function of $\Phi$. We accordingly  find \begin{widetext}
\beq \label{I}
G_{\rm tun} \propto |\lambda|^2 k_{\rm F} W e^2  \int_{1/N}^{1-1/N} \frac{dq}{1-q^2} \left\{1+q^2+ \cos \Phi \cos\left[ k_{\rm F}(\Delta x^{\rm u} \! -\Delta x^{\rm l}) \sqrt{1-q^2}\right] \right\}
\eeq
\end{widetext}
 for the differential tunnel  conductance $G_{\rm tun}= dI_{\rm tun}/d\delta V|_{\delta V=0}$ at   generic $\varepsilon_{\rm F}$ and  $| \Delta x^{\rm u}-\Delta x^{\rm l}|\ll \Delta x^{\rm u}$ \cite{unpub}.
Here $1/N \sim 1/k_{\rm F}W$  depends on the Fermi wavevector of the lower ribbon $k_{\rm F}$. We conclude that in the geometry of Fig.\ \ref{fig1} the predicted exchange vector potential $\boldsymbol{A^{\rm ex}}$ has the very distinctive signature that the tunnel conductance $G_{\rm tun}$ {\em between} the two ribbons oscillates as a function of   the current  flowing {\em through} the upper ribbon. 


  The visibility of the predicted oscillations is quantified by the  interference contrast $C = G_{\rm int}/G_{\rm dir}$.  We extract  the `direct'  conductance $G_{\rm dir}$ and the oscillatory  `interference' conductance $G_{\rm int}$  from Eq.\ (\ref{I}) as the
  first two and the third terms in curly brackets, respectively.
   We   find  $C\sim 1$    for  $k_{\rm F}\Delta x^\mu \gg \sqrt{N}$ ($\mu \in\{{\rm u, l}\}$), but  $k_{\rm F}|\Delta x^{\rm u}-\Delta x^{\rm l}| \ll \sqrt{N}$.  Larger path length differences   $k_{\rm F}\Delta x^\mu \gg  k_{\rm F}|\Delta x^{\rm u}-\Delta x^{\rm l}| \gg \sqrt{N}$ suppress the interference contrast,  $C\sim  \sqrt{N}/ k_{\rm F}|\Delta x^{\rm u} \!-\Delta x^{\rm l} |\ln N$. Above we assumed tunneling matrix elements $\lambda$ that are constant over the width of the ribbons, so that electrons tunnel only between identical transverse modes.   To estimate the influence of inhomogeneities of $\lambda$  we compute the tunneling conductance  also for   random tunnel matrix elements with identical ensemble averages for scattering between any two of the  transverse modes. We find $C\sim 1/ N $ and we expect that $C$ suffers similar or weaker suppressions for most realistic forms of tunnel contacts.

 Violations of our limit $\Delta x^\mu \lesssim l_{\rm in}$  suppress $C$ exponentially.   According to Eq.\ (\ref{lin}) the condition $\Delta x^\mu \lesssim l_{\rm in}$  is met at low voltages. Demanding  at the same time that $V$  be large enough to produce  one entire oscillation of $G_{\rm tun}$  imposes a lower bound on the path lengths  
   $ k_{\rm F} \Delta x^\mu\gtrsim 32 \pi^2 r^{1/2}_{\rm s}\zeta^{-3/2} $ \cite{footnote3}. Note that Eq.\ (\ref{lin}) assumes a two-dimensional electron gas, $W \gg v/eV$. In the one-dimensional limit, realizable in carbon nanotubes, quasi-particles with well-defined energies cease to exist and the above predictions do not apply.
  Inelastic or elastic scattering that induces a potential drop $\Delta V$ along the ribbons may also cause an electric Aharonov-Bohm effect. In order for the resulting oscillations not to obscure the predicted effect it has to be assured that     $\Delta V \ll v/e \Delta x^\mu$.

\begin{figure}
\includegraphics[width=5cm]{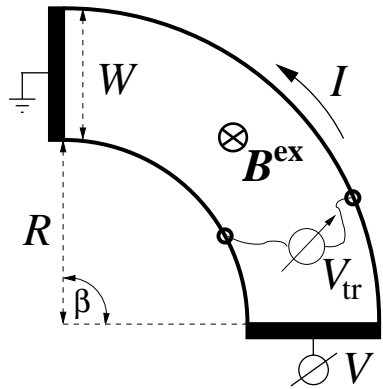}
\caption{ Graphene annulus of width $W$ and inner radius $R$, contacted by electron reservoirs (solid black bars) whose electrochemical potentials differ by $eV$. A current $I$ is flowing that produces a transverse voltage $V_{\rm tr}$ by a Hall effect in the (fictitious) exchange magnetic field ${\bf B}^{\rm ex}$.  }  \label{fig2}
\end{figure}

{\it Hall effect:} In a curved conductor, cut out of a sheet of graphene, the current direction $\boldsymbol{w_\theta}$ is not constant anymore. We thus expect an exchange vector potential $\boldsymbol{A^{\rm ex}}$ similar to Eq.\ (\ref{Aex}), but space-dependent. The  effective magnetic field that would generically result should be observable by the Hall effect that it causes. To confirm this scenario we consider a segment of a graphene annulus with inner radius $R$, radial width $W$, and angular width $\beta$  (see Fig.\ \ref{fig2}). We again address the semiclassical limit  $k_{\rm F} W \gg 1$ \cite{footnote1} and  omit the details of the calculation that will be presented elsewhere \cite{unpub}.  The segment is well-coupled to electron reservoirs at its angular boundaries, such that electrical current  flows in the angular direction  and we have periodic boundary conditions at the angular edges.
  The wavefunctions   with these boundary conditions obtain   from the solutions of Eq.\ (\ref{Dirac}) in the full annulus, where angular momentum $l$ is a good quantum number. Additionally the single-particle  states $\psi_{\sigma \gamma sn l}$   are  characterized by a radial  quantum number $n$   and they have eigenenergies $\varepsilon_{snl}$.
 At $r_{\rm s}=0$ a voltage bias between the two reservoirs in Fig.\ \ref{fig2} creates an occupation of states $f_{  snl} =   \Theta[\varepsilon_{\rm F}+\Theta(sl) eV-\varepsilon_{snl}] $   ($T=0$).
 Substituting the corresponding $G^{(0)K}$ into   Eq.\ (\ref{G0W}),  in our `screened Hartree-Fock' approximation corresponding to Eq.\ (\ref{Fock}),   and extracting $\boldsymbol{A^{\rm ex}}$ through Eq.\  (\ref{project}) yields an exchange vector potential   $\boldsymbol{A^{\rm ex}}(r\boldsymbol{\hat{w}_{\varphi}})=\zeta(c/v) (V/8\pi) {\boldsymbol{\hat{w}_{\varphi+\pi/2}}}$ as suggested by Eq.\ (\ref{Aex}), pointing
 in angular direction.

 The above $\boldsymbol{A^{\rm ex}}$ implies an effective magnetic field   $B^{\rm ex}=|\boldsymbol{A^{\rm ex}}|/r$ perpendicular to the annulus. The corresponding Lorentz force modifies the charge density $\rho$, found from the Green functions implied by  $\Sigma^{(0)}$,
Eq.\ (\ref{Fock}),
  \beq \label{rho}
  \rho(r)=\sum_{\sigma \gamma snl}  f^{\rm ex}_{snl} |\psi^{\rm ex}_{\sigma\gamma snl}(r)|^2 .
  \eeq
 Here, $\psi^{\rm ex}_{\sigma\gamma snl}$ and  $f^{\rm ex}_{snl}$ follow from Eq.\ (\ref{Dirac}) with $\boldsymbol{ A}=\boldsymbol{A^{\rm ex}}$ 
 and in the semiclassical limit we obtain
   \beq \label{rho0}
 \rho(r)=\rho_0-\frac{1}{2\pi} \left(\frac{eV}{2\pi  v}\right)^2\left[\frac{R}{r}\frac{R+W}{W} \ln\left(1+ \frac{W}{R}\right)-1\right] .
 \eeq
  The equilibrium contribution $\rho_0$ is $r$-independent \cite{footnote4}. The     $r$-dependent contribution to $\rho$, Eq.\ (\ref{rho0}),  is screened by a charge rearrangement through   a transverse voltage
  \beq \label{Vtr}
 V_{\rm tr}^{\rm ex}= \frac{V}{8\pi^2} \frac{eV}{\varepsilon_{\rm F}} \ln\left(1+\frac{W}{R}\right) +{\cal O}(e^3V^3/\varepsilon_{\rm F}^2),
 \eeq
   restoring charge neutrality. Eq.\ (\ref{Vtr}) predicts a transverse `Hall' voltage generated by an electrical current in the absence of a real magnetic field. It is a second peculiar signature of the  exchange vector potential $\boldsymbol{A^{\rm ex}}$.

 Eq.\ (\ref{Vtr}) crucially relies on our assumption that the annulus is ballistic,   $l_{\rm m} \gg L$, where  $L={\rm max}\{\beta R, W\}$ and $l_{\rm m}={\rm min}\{ l_{\rm el}, \, l_{\rm in}\}$.
 In the opposite limit of a diffusive annulus, $L \gg l_{\rm m} $, the current density decays as $1/r$, implying   $\boldsymbol{A^{\rm ex}} \propto \boldsymbol{\hat{w}_{\varphi+\pi/2}}/r$ with ${B}^{\rm ex}=0$ and thus $V_{\rm tr}^{\rm ex}=0$. With moderately strong backscattering, $l_{\rm m} \sim L$,  $\boldsymbol{A^{\rm ex}}$ still produces a transverse voltage.  Backscattering, however,  induces a voltage drop along the annulus that generically produces an additional transverse voltage $V_{\rm tr}^{\rm scatt}$. The linear in $V$ contribution to   $V_{\rm tr}^{\rm scatt}$ can be eliminated by measuring $\bar{V}_{\rm tr}= [V_{\rm tr}(V)+V_{\rm tr}(-V)]$.  Any energy-dependence of the scattering, however, will add a quadratic in $V$ contribution to $V^{\rm scatt}_{\rm tr}$ that competes with $V_{\rm tr}^{\rm ex}$. Typically backscattering in graphene has an energy-dependence on the scale $\varepsilon_{\rm F}$ \cite{nomura:prl07,kumazaki:jpj06,ando:jpj06,hwang:prl07}. It follows that $\bar{V}^{\rm scatt}_{\rm tr} $ is negligible, $\bar{V}^{\rm scatt}_{\rm tr}\ll \bar{V}_{\rm tr}^{\rm ex}$, if $ L/l_{\rm m} \ll \ln(1+W/R)$. Experimentally   $\bar{V}^{\rm scatt}_{\rm tr}$ can be distinguished from $V_{\rm tr}^{\rm ex}$  by varying   $\varepsilon_{\rm F}$ at $eV\ll \varepsilon_{\rm F}$. If the measured $\bar{V}_{\rm tr}$ is due to energy-dependent scattering, a variation of $\varepsilon_{\rm F}$ by $d\mu\simeq eV$ causes a relative change of $\bar{V}_{\rm tr}$ by $\Delta \bar{V}_{\rm tr}^{\rm scatt} /\bar{V}_{\rm tr}^{\rm scatt} = ( \bar{V}_{\rm tr}^{\rm scatt}(V)|_{\varepsilon_{\rm F}+d\mu}-\bar{V}_{\rm tr}^{\rm scatt}(V)|_{\varepsilon_{\rm F}} )/\bar{V}_{\rm tr}^{\rm scatt}(V)|_{\varepsilon_{\rm F}} \simeq 1$. In contrast, the relative change is  only $\Delta \bar{V}_{\rm tr}^{\rm ex}/\bar{V}_{\rm tr}^{\rm ex} \simeq d\mu/\varepsilon_{\rm F}\ll 1$ if ${V}_{\rm tr}$ is due to   the exchange interaction.

 A second effect competing with $V^{\rm ex}_{\rm tr}$ is the Hall effect in the (real) magnetic field produced by the current flowing in the annulus. In the limit $W/R \ll 1$ the relative magnitude of the resulting transverse voltage $ \bar{V}_{\rm tr}^{\rm magn}$ is
 \beq
 \frac{ \bar{V}_{\rm tr}^{\rm magn}}{\bar{V}_{\rm tr}^{\rm ex}} = \frac{8}{\zeta\pi}\left(\frac {v}{c}\right)^2 \frac{e^2}{v } \left[\ln\frac{8R \tan (\beta/4)}{W})+\frac{3}{2}\right] k_{\rm F} W
  \eeq
  up to corrections of order $(W/R)\ln(W/R)$. For the parameters $W=0.2 R$, $\beta=\pi/6$, and $r_{\rm s}=0.15$ we find for instance $ \bar{V}_{\rm tr}^{\rm magn}/ \bar{V}_{\rm tr}^{\rm ex} \approx 4 \cdot 10^{-4}\, k_{\rm F} W$. For typical sample dimensions of $W \approx 1\mu {\rm m}$ this effect is thus negligible up to the largest $k_{\rm F} \approx 1 \,{\rm nm}$. Also the smallness of $ \bar{V}_{\rm tr}^{\rm magn}$ can be verified experimentally by its distinctive dependence on $\varepsilon_{\rm F}$. In this case $\bar{V}_{\rm tr}^{\rm magn} $ is independent of $\varepsilon_{\rm F}$,   $\Delta \bar{V}_{\rm tr}^{\rm magn}/ \bar{V}_{\rm tr}^{\rm magn} = 0$, compared to $\Delta \bar{V}_{\rm tr}^{\rm ex}/ \bar{V}_{\rm tr}^{\rm ex} \simeq d\mu/\varepsilon_{\rm F}$. Both  competing effects discussed above can be ruled out by simultaneously varying $\varepsilon_{\rm F}$ and $V$.

 {\it Conclusions:} We have predicted a novel nonequilibrium effect in coherent two-dimensional electronic systems. The effect is due to electron-electron interactions and it generates a fictitious vector potential that  has striking signatures: interference currents that oscillate as a function of the current density in the material and a `Hall' voltage  in zero magnetic field.    We have made detailed predictions for graphene, where the effect is particularly robust owing to the  pseudospin  degree of freedom of the conduction electrons.

 The author thanks C.\ W.\ J.\ Beenakker, P.\ W.\ Brouwer, R.\ Mani, Y.\ V.\ Nazarov, and A.\ Zangwill for discussions.


\vspace{-.4cm}


\begin{thebibliography}{30}
\vspace{-.4cm}
\expandafter\ifx\csname natexlab\endcsname\relax\def\natexlab#1{#1}\fi
\expandafter\ifx\csname bibnamefont\endcsname\relax
  \def\bibnamefont#1{#1}\fi
\expandafter\ifx\csname bibfnamefont\endcsname\relax
  \def\bibfnamefont#1{#1}\fi
\expandafter\ifx\csname citenamefont\endcsname\relax
  \def\citenamefont#1{#1}\fi
\expandafter\ifx\csname url\endcsname\relax
  \def\url#1{\texttt{#1}}\fi
\expandafter\ifx\csname urlprefix\endcsname\relax\def\urlprefix{URL }\fi
\providecommand{\bibinfo}[2]{#2}
\providecommand{\eprint}[2][]{\url{#2}}

\bibitem[{\citenamefont{Ji et~al.}(2003)\citenamefont{Ji, Chung, Sprinzak,
  Heiblum, Mahalu, and Shtrikman}}]{Ji:nat03}
\bibinfo{author}{\bibfnamefont{Y.}~\bibnamefont{Ji}},
  \bibinfo{author}{\bibfnamefont{Y.}~\bibnamefont{Chung}},
  \bibinfo{author}{\bibfnamefont{D.}~\bibnamefont{Sprinzak}},
  \bibinfo{author}{\bibfnamefont{M.}~\bibnamefont{Heiblum}},
  \bibinfo{author}{\bibfnamefont{D.}~\bibnamefont{Mahalu}}, \bibnamefont{and}
  \bibinfo{author}{\bibfnamefont{H.}~\bibnamefont{Shtrikman}},
  \bibinfo{journal}{Nature} \textbf{\bibinfo{volume}{422}},
  \bibinfo{pages}{415} (\bibinfo{year}{2003}).

\bibitem[{\citenamefont{Goldhaber-Gordon
  et~al.}(1998)\citenamefont{Goldhaber-Gordon, Shtrikman, Mahalu,
  Abusch-Magder, Meirav, and Kastner}}]{goldhaber:nat98}
\bibinfo{author}{\bibfnamefont{D.}~\bibnamefont{Goldhaber-Gordon}},
  \bibinfo{author}{\bibfnamefont{H.}~\bibnamefont{Shtrikman}},
  \bibinfo{author}{\bibfnamefont{D.}~\bibnamefont{Mahalu}},
  \bibinfo{author}{\bibfnamefont{D.}~\bibnamefont{Abusch-Magder}},
  \bibinfo{author}{\bibfnamefont{U.}~\bibnamefont{Meirav}}, \bibnamefont{and}
  \bibinfo{author}{\bibfnamefont{M.~A.} \bibnamefont{Kastner}},
  \bibinfo{journal}{Nature} \textbf{\bibinfo{volume}{{\bf 391}}},
  \bibinfo{pages}{156} (\bibinfo{year}{1998}).

\bibitem[{\citenamefont{Cronenwett et~al.}(1998)\citenamefont{Cronenwett,
  Oosterkamp, and Kouwenhoven}}]{cronenwett:sci98}
\bibinfo{author}{\bibfnamefont{S.~M.} \bibnamefont{Cronenwett}},
  \bibinfo{author}{\bibfnamefont{T.~H.} \bibnamefont{Oosterkamp}},
  \bibnamefont{and} \bibinfo{author}{\bibfnamefont{L.~P.}
  \bibnamefont{Kouwenhoven}}, \bibinfo{journal}{Science}
  \textbf{\bibinfo{volume}{281}}, \bibinfo{pages}{540} (\bibinfo{year}{1998}).

\bibitem[{\citenamefont{Josephson}(1974)}]{josephson:rmp74}
\bibinfo{author}{\bibfnamefont{B.~D.} \bibnamefont{Josephson}},
  \bibinfo{journal}{Rev. Mod. Phys.} \textbf{\bibinfo{volume}{46}},
  \bibinfo{pages}{251} (\bibinfo{year}{1974}).

\bibitem[{\citenamefont{Novoselov et~al.}(2004)\citenamefont{Novoselov, Geim,
  Morozov, Jiang, Zhang, Dubonos, Grigorieva, and Firsov}}]{novoselov:sci04}
\bibinfo{author}{\bibfnamefont{K.}~\bibnamefont{Novoselov}},
  \bibinfo{author}{\bibfnamefont{A.}~\bibnamefont{Geim}},
  \bibinfo{author}{\bibfnamefont{S.}~\bibnamefont{Morozov}},
  \bibinfo{author}{\bibfnamefont{D.}~\bibnamefont{Jiang}},
  \bibinfo{author}{\bibfnamefont{Y.}~\bibnamefont{Zhang}},
  \bibinfo{author}{\bibfnamefont{S.}~\bibnamefont{Dubonos}},
  \bibinfo{author}{\bibfnamefont{I.}~\bibnamefont{Grigorieva}},
  \bibnamefont{and} \bibinfo{author}{\bibfnamefont{A.}~\bibnamefont{Firsov}},
  \bibinfo{journal}{Science} \textbf{\bibinfo{volume}{306}},
  \bibinfo{pages}{666} (\bibinfo{year}{2004}).

\bibitem[{\citenamefont{Zhang et~al.}(2005)\citenamefont{Zhang, Tan, Stormer,
  and Kim}}]{zhang:nat05}
\bibinfo{author}{\bibfnamefont{Y.}~\bibnamefont{Zhang}},
  \bibinfo{author}{\bibfnamefont{Y.-W.} \bibnamefont{Tan}},
  \bibinfo{author}{\bibfnamefont{H.~L.} \bibnamefont{Stormer}},
  \bibnamefont{and} \bibinfo{author}{\bibfnamefont{P.}~\bibnamefont{Kim}},
  \bibinfo{journal}{Nature} \textbf{\bibinfo{volume}{438}},
  \bibinfo{pages}{201} (\bibinfo{year}{2005}).

\bibitem[{\citenamefont{Berger et~al.}(2004)\citenamefont{Berger, Song, Li, Li,
  Ogbazghi, Feng, Dai, Marchenkov, Conrad, First et~al.}}]{berger:jpc04}
\bibinfo{author}{\bibfnamefont{C.}~\bibnamefont{Berger}},
  \bibinfo{author}{\bibfnamefont{Z.}~\bibnamefont{Song}},
  \bibinfo{author}{\bibfnamefont{T.}~\bibnamefont{Li}},
  \bibinfo{author}{\bibfnamefont{X.}~\bibnamefont{Li}},
  \bibinfo{author}{\bibfnamefont{A.~Y.} \bibnamefont{Ogbazghi}},
  \bibinfo{author}{\bibfnamefont{R.}~\bibnamefont{Feng}},
  \bibinfo{author}{\bibfnamefont{Z.}~\bibnamefont{Dai}},
  \bibinfo{author}{\bibfnamefont{A.~N.} \bibnamefont{Marchenkov}},
  \bibinfo{author}{\bibfnamefont{E.~H.} \bibnamefont{Conrad}},
  \bibinfo{author}{\bibfnamefont{P.~N.} \bibnamefont{First}},
  \bibnamefont{et~al.}, \bibinfo{journal}{J. Phys. Chem. B}
  \textbf{\bibinfo{volume}{108}}, \bibinfo{pages}{19912}
  (\bibinfo{year}{2004}).

\bibitem[{\citenamefont{Geim and Novoselov}(2007)}]{novoselov:nma07}
\bibinfo{author}{\bibfnamefont{A.~K.} \bibnamefont{Geim}} \bibnamefont{and}
  \bibinfo{author}{\bibfnamefont{K.~S.} \bibnamefont{Novoselov}},
  \bibinfo{journal}{Nature Materials} \textbf{\bibinfo{volume}{6}},
  \bibinfo{pages}{183} (\bibinfo{year}{2007}).

\bibitem[{\citenamefont{Katsnelson et~al.}(2006)\citenamefont{Katsnelson,
  Novoselov, and Geim}}]{katsnelson:nph06}
\bibinfo{author}{\bibfnamefont{M.~I.} \bibnamefont{Katsnelson}},
  \bibinfo{author}{\bibfnamefont{K.~S.} \bibnamefont{Novoselov}},
  \bibnamefont{and} \bibinfo{author}{\bibfnamefont{A.~K.} \bibnamefont{Geim}},
  \bibinfo{journal}{Nature Phys.} \textbf{\bibinfo{volume}{2}},
  \bibinfo{pages}{620} (\bibinfo{year}{2006}).

\bibitem[{\citenamefont{Barlas et~al.}(2007)\citenamefont{Barlas, Pereg-Barnea,
  Polini, Asgari, and MacDonald}}]{barlas:prl07}
\bibinfo{author}{\bibfnamefont{Y.}~\bibnamefont{Barlas}},
  \bibinfo{author}{\bibfnamefont{T.}~\bibnamefont{Pereg-Barnea}},
  \bibinfo{author}{\bibfnamefont{M.}~\bibnamefont{Polini}},
  \bibinfo{author}{\bibfnamefont{R.}~\bibnamefont{Asgari}}, \bibnamefont{and}
  \bibinfo{author}{\bibfnamefont{A.~H.} \bibnamefont{MacDonald}},
  \bibinfo{journal}{Phys. Rev. Lett.} \textbf{\bibinfo{volume}{98}},
  \bibinfo{eid}{236601} (\bibinfo{year}{2007}).

\bibitem[{\citenamefont{Iordanskii and Koshelev}(1985)}]{iordanskii:jet85}
\bibinfo{author}{\bibfnamefont{S.}~\bibnamefont{Iordanskii}} \bibnamefont{and}
  \bibinfo{author}{\bibfnamefont{A.}~\bibnamefont{Koshelev}},
  \bibinfo{journal}{JETP Letters} \textbf{\bibinfo{volume}{41}},
  \bibinfo{pages}{574} (\bibinfo{year}{1985}).

\bibitem[{\citenamefont{Morozov et~al.}(2006)\citenamefont{Morozov, Novoselov,
  Katsnelson, Schedin, Ponomarenko, Jiang, and Geim}}]{morozov:prl06}
\bibinfo{author}{\bibfnamefont{S.~V.} \bibnamefont{Morozov}},
  \bibinfo{author}{\bibfnamefont{K.~S.} \bibnamefont{Novoselov}},
  \bibinfo{author}{\bibfnamefont{M.~I.} \bibnamefont{Katsnelson}},
  \bibinfo{author}{\bibfnamefont{F.}~\bibnamefont{Schedin}},
  \bibinfo{author}{\bibfnamefont{L.~A.} \bibnamefont{Ponomarenko}},
  \bibinfo{author}{\bibfnamefont{D.}~\bibnamefont{Jiang}}, \bibnamefont{and}
  \bibinfo{author}{\bibfnamefont{A.~K.} \bibnamefont{Geim}},
  \bibinfo{journal}{Phys. Rev. Lett.} \textbf{\bibinfo{volume}{97}},
  \bibinfo{eid}{016801} (\bibinfo{year}{2006}).

\bibitem[{\citenamefont{Morpurgo and Guinea}(2006)}]{morpurgo:prl06}
\bibinfo{author}{\bibfnamefont{A.~F.} \bibnamefont{Morpurgo}} \bibnamefont{and}
  \bibinfo{author}{\bibfnamefont{F.}~\bibnamefont{Guinea}},
  \bibinfo{journal}{Phys. Rev. Lett.} \textbf{\bibinfo{volume}{97}},
  \bibinfo{eid}{196804} (\bibinfo{year}{2006}).

\bibitem[{\citenamefont{Rammer and Smith}(1986)}]{Rammer:rmp86}
\bibinfo{author}{\bibfnamefont{J.}~\bibnamefont{Rammer}} \bibnamefont{and}
  \bibinfo{author}{\bibfnamefont{H.}~\bibnamefont{Smith}},
  \bibinfo{journal}{Rev. Mod. Phys.} \textbf{\bibinfo{volume}{58}},
  \bibinfo{pages}{323} (\bibinfo{year}{1986}).

\bibitem[{\citenamefont{Hwang and Sarma}(2007)}]{hwang:prb07}
\bibinfo{author}{\bibfnamefont{E.~H.} \bibnamefont{Hwang}} \bibnamefont{and}
  \bibinfo{author}{\bibfnamefont{S.~D.} \bibnamefont{Sarma}},
  \bibinfo{journal}{Phys. Rev. B} \textbf{\bibinfo{volume}{75}},
  \bibinfo{eid}{205418} (\bibinfo{year}{2007}).

\bibitem[{\citenamefont{Wunsch et~al.}(2006)\citenamefont{Wunsch, Stauber,
  Sols, and Guinea}}]{wunsch:njp06}
\bibinfo{author}{\bibfnamefont{B.}~\bibnamefont{Wunsch}},
  \bibinfo{author}{\bibfnamefont{T.}~\bibnamefont{Stauber}},
  \bibinfo{author}{\bibfnamefont{F.}~\bibnamefont{Sols}}, \bibnamefont{and}
  \bibinfo{author}{\bibfnamefont{F.}~\bibnamefont{Guinea}},
  \bibinfo{journal}{New J. Phys.} \textbf{\bibinfo{volume}{8}},
  \bibinfo{pages}{318} (\bibinfo{year}{2006}).

\bibitem[{\citenamefont{Das~Sarma et~al.}(2007)\citenamefont{Das~Sarma, Hu,
  Hwang, and Tse}}]{DasSarma:07}
\bibinfo{author}{\bibfnamefont{S.}~\bibnamefont{Das~Sarma}},
  \bibinfo{author}{\bibfnamefont{B.~Y.-K.} \bibnamefont{Hu}},
  \bibinfo{author}{\bibfnamefont{E.~H.} \bibnamefont{Hwang}}, \bibnamefont{and}
  \bibinfo{author}{\bibfnamefont{W.-K.} \bibnamefont{Tse}},
  \bibinfo{journal}{arXiv:0708.3239}  (\bibinfo{year}{2007}).

\bibitem[{foo({\natexlab{a}})}]{footnote11}
\bibinfo{note}{This is expected since the Hamiltonian of interacting electrons
  in graphene at $\boldsymbol{A}=0$ can be brought into a valley isotropic form
  by a unitary transformation \cite{akhmerov:prl07}.}

\bibitem[{foo({\natexlab{b}})}]{footnote2}
\bibinfo{note}{A numerical analysis that neglects the frequency-dependence of
  $W^{\rm R}$ results in corrections to $\zeta$ of relative magnitude $\approx
  20\% $ at $r_{\rm s}=0.15$ and suggests that they are negligible in weakly
  interacting graphene samples.}

\bibitem[{\citenamefont{Kindermann}()}]{unpub}
\bibinfo{author}{\bibfnamefont{M.}~\bibnamefont{Kindermann}},
  \bibinfo{journal}{to be published}.

\bibitem[{\citenamefont{Sarma and Hwang}(2005)}]{Sarma:prb05}
\bibinfo{author}{\bibfnamefont{S.~D.} \bibnamefont{Sarma}} \bibnamefont{and}
  \bibinfo{author}{\bibfnamefont{E.~H.} \bibnamefont{Hwang}},
  \bibinfo{journal}{Phys. Rev. B} \textbf{\bibinfo{volume}{72}},
  \bibinfo{eid}{035311} (\bibinfo{year}{2005}).

\bibitem[{foo({\natexlab{c}})}]{footnote1}
\bibinfo{note}{Our results obtain for metallic armchair, semiconducting
  armchair or zigzag edges, suggesting that they do not depend sensitively on
  the boundary conditions provided there is no backscattering from boundary
  roughness.}

\bibitem[{foo({\natexlab{d}})}]{footnote3}
\bibinfo{note}{An electric Aharonov-Bohm effect through inelastic
  backscattering in the upper ribbon is ruled out under a similar condition.}

\bibitem[{foo({\natexlab{e}})}]{footnote4}
\bibinfo{note}{In the limit $k_{\rm F} W \sim 1$ transverse electric fields
  appear also in equilibrium \cite{simanek:pla98}.}

\bibitem[{\citenamefont{Nomura and MacDonald}(2007)}]{nomura:prl07}
\bibinfo{author}{\bibfnamefont{K.}~\bibnamefont{Nomura}} \bibnamefont{and}
  \bibinfo{author}{\bibfnamefont{A.~H.} \bibnamefont{MacDonald}},
  \bibinfo{journal}{Phys. Rev. Lett.} \textbf{\bibinfo{volume}{98}},
  \bibinfo{eid}{076602} (\bibinfo{year}{2007}).

\bibitem[{\citenamefont{Kumazaki and Hirashima}(2006)}]{kumazaki:jpj06}
\bibinfo{author}{\bibfnamefont{H.}~\bibnamefont{Kumazaki}} \bibnamefont{and}
  \bibinfo{author}{\bibfnamefont{D.}~\bibnamefont{Hirashima}},
  \bibinfo{journal}{J. Phys. Soc. Jpn.} \textbf{\bibinfo{volume}{75}},
  \bibinfo{pages}{053707} (\bibinfo{year}{2006}).

\bibitem[{\citenamefont{Ando}(2006)}]{ando:jpj06}
\bibinfo{author}{\bibfnamefont{T.}~\bibnamefont{Ando}}, \bibinfo{journal}{J.
  Phys. Soc. Jpn.} \textbf{\bibinfo{volume}{75}}, \bibinfo{pages}{074716}
  (\bibinfo{year}{2006}).

\bibitem[{\citenamefont{Hwang et~al.}(2007)\citenamefont{Hwang, Adam, and
  Sarma}}]{hwang:prl07}
\bibinfo{author}{\bibfnamefont{E.~H.} \bibnamefont{Hwang}},
  \bibinfo{author}{\bibfnamefont{S.}~\bibnamefont{Adam}}, \bibnamefont{and}
  \bibinfo{author}{\bibfnamefont{S.~D.} \bibnamefont{Sarma}},
  \bibinfo{journal}{Phys. Rev. Lett.} \textbf{\bibinfo{volume}{98}},
  \bibinfo{eid}{186806} (\bibinfo{year}{2007}).

\bibitem[{\citenamefont{Akhmerov and Beenakker}(2007)}]{akhmerov:prl07}
\bibinfo{author}{\bibfnamefont{A.~R.} \bibnamefont{Akhmerov}} \bibnamefont{and}
  \bibinfo{author}{\bibfnamefont{C.~W.~J.} \bibnamefont{Beenakker}},
  \bibinfo{journal}{Phys. Rev. Lett.} \textbf{\bibinfo{volume}{98}},
  \bibinfo{eid}{157003} (\bibinfo{year}{2007}).

\bibitem[{\citenamefont{Simanek}(1998)}]{simanek:pla98}
\bibinfo{author}{\bibfnamefont{E.}~\bibnamefont{Simanek}},
  \bibinfo{journal}{Phys. Lett. A} \textbf{\bibinfo{volume}{250}},
  \bibinfo{pages}{425} (\bibinfo{year}{1998}).

\end{thebibliography}
 \end{document}